\documentclass[10pt,conference]{IEEEtran}
\IEEEoverridecommandlockouts
\usepackage{cite}
\usepackage{amsmath,amssymb,amsfonts}
\usepackage{algorithm}
\usepackage{algorithmicx}
\usepackage{algpseudocode}
\usepackage{enumitem}
\usepackage{multirow}
\usepackage{supertabular}
\usepackage{graphicx}
\usepackage{textcomp}
\usepackage{xcolor}
\usepackage{booktabs}
\def\BibTeX{{\rm B\kern-.05em{\sc i\kern-.025em b}\kern-.08em
    T\kern-.1667em\lower.7ex\hbox{E}\kern-.125emX}}

\usepackage{float,subcaption}
\usepackage{tikz,tikzscale}
\usepackage{pgfplots,lipsum}
\usepgfplotslibrary{statistics}

\makeatletter
\newtoks\therules
\therules={}
\def\appendto#1#2{\expandafter#1\expandafter{\the#1#2}}
\def\gobblefirst#1{
  #1\expandafter\expandafter\expandafter{\expandafter\@gobble\the#1}}%
\def\LState{\State\unskip\the\therules}
\def\printindent{\unskip\the\therules}%
\begin{document}

\title{
Advancing Federated Learning in 6G: A Trusted Architecture with Graph-based Analysis
}

\author{\IEEEauthorblockN{Wenxuan Ye\IEEEauthorrefmark{1}\IEEEauthorrefmark{2},
Chendi Qian\IEEEauthorrefmark{3},
Xueli An\IEEEauthorrefmark{1}, Xueqiang Yan\IEEEauthorrefmark{4},
Georg Carle\IEEEauthorrefmark{2}}
\IEEEauthorblockA{
\IEEEauthorrefmark{1} Advanced Wireless Technology Laboratory, Munich Research Center, Huawei Technologies Duesseldorf GmbH\\
\IEEEauthorrefmark{2} TUM School of Computation, Information and Technology, Technical University of Munich\\
\IEEEauthorrefmark{3} Computer Science 6: Machine Learning and Reasoning, RWTH Aachen\\
\IEEEauthorrefmark{4} Wireless Technology Lab, 2012 Laboratories, Huawei Technologies Co., Ltd\\
wenxuan.ye@tum.de, chendi.qian@log.rwth-aachen.de, \{xueli.an, yanxueqiang1\}@huawei.com, carle@net.in.tum.de
}}

\maketitle

\begin{abstract}
Integrating native AI support into the network architecture is an essential objective of 6G.
Federated Learning (FL) emerges as a potential paradigm, facilitating decentralized AI model training across a diverse range of devices under the coordination of a central server.
However, several challenges hinder its wide application in the 6G context, such as malicious attacks and privacy snooping on local model updates, and centralization pitfalls.
This work proposes a trusted architecture for supporting FL, which utilizes Distributed Ledger Technology (DLT) and Graph Neural Network (GNN), including three key features.
First, a pre-processing layer employing homomorphic encryption is incorporated to securely aggregate local models, preserving the privacy of individual models.
Second, given the distributed nature and graph structure between clients and nodes in the pre-processing layer, GNN is leveraged to identify abnormal local models, enhancing system security.
Third, DLT is utilized to decentralize the system by selecting one of the candidates to perform the central server's functions. 
Additionally, DLT ensures reliable data management by recording data exchanges in an immutable and transparent ledger.
The feasibility of the novel architecture is validated through simulations, demonstrating improved performance in anomalous model detection and global model accuracy compared to relevant baselines.
\end{abstract}

\begin{IEEEkeywords}
Federated learning, Distributed ledger technology, Graph neural network, Secure aggregation, 6G, Homomorphic encryption 
\end{IEEEkeywords}

\algblock{ParFor}{EndParFor}
\algnewcommand\algorithmicparfor{\textbf{parfor}}
\algnewcommand\algorithmicpardo{\textbf{do}}
\algnewcommand\algorithmicendparfor{\textbf{end\ parfor}}
\algrenewtext{ParFor}[1]{\algorithmicparfor\ #1\ \algorithmicpardo}
\algrenewtext{EndParFor}{\algorithmicendparfor}

\section{INTRODUCTION}
\label{sec:introduction}

6G regards AI as a crucial enabler driving innovation in network architecture.
In addition to utilizing AI for intelligent communication networks, 6G aims to achieve native AI support, transcending the traditional role of delivery channels for collected data \cite{Tong2021}.
Federated Learning (FL), a distributed learning paradigm with efficient communication, emerges as a promising solution for extracting valuable information from dispersed client data while ensuring privacy \cite{McMahan2017}.
A central server distributes the training process across multiple clients, with each client performing local computations on its own data, and iteratively exchanging model parameters to improve model quality.
It holds diverse potentials across vertical industries, including smart cities, healthcare and other domains involving extensive data exchanges and analysis \cite{Tong2021}.
Recent advancements in decentralized computing and communication capabilities of mobile networks have further facilitated the provision of FL services \cite{Liu2020}. 

However, several challenges hinder the wide application of FL in the 6G context.
From a \textit{security} perspective, detecting and addressing anomalous models pose notable obstacles due to the restriction of analyzing only the model updates without direct access to the raw data \cite{Wang2019}.
Adversarial clients may launch poisoning attacks to impair the model performance and disrupt the training process.
From a \textit{privacy} view, while FL offers a considerable improvement over centralized training, it is still susceptible to information leakage through the observation of clients' shared model updates \cite{Kairouz2021}.
From an \textit{architectural} standpoint, with massive connectivity, the central server may struggle to handle the increased load.
The centralized design for accessing and processing model parameters is prone to a single point of failure and limited scalability. 
Besides, such a centralized system lacks a proactive mechanism for demonstrating compliance in data management \cite{Tong2021}.

To enhance privacy preservation in FL, secure aggregation has been recognized as a pivotal technology, enabling the combination of model updates without disclosing individual contributions \cite{Bonawitz2017}.
Homomorphic Encryption (HE) is a promising cryptographic method for implementing secure aggregation \cite{Gilad2016}. 
It allows the server to perform computations directly on encrypted updates to obtain the desired encrypted result.
Recent progress in its utility, coupled with increased computing capabilities of communication network entities, have paved the way for its widespread application \cite{Gilad2016}.

In addition to secure aggregation for privacy issues, Distributed Ledger Technology (DLT) offers a potential solution to overcome other FL limitations, such as centralization issues and unreliable data management.
DLT ensures transparency and immutability by decentralized peers maintaining a shared ledger for blocks \cite{Nakamoto2009}.
Each block stores multiple transactions and links to the previous block by containing its cryptographic hash, forming a chronological record \cite{Androulaki2018}.
The smart contract, a piece of code deployed on a DLT platform, executes automatically when preset conditions are met \cite{Androulaki2018}.
It reduces the need for intermediaries and streamlines processes.

Furthermore, considering the complex relationships among data and the distributed nature of FL, Graph Neural Networks (GNNs) have emerged as a powerful tool for modeling graph-structured data \cite{Scarselli2008}.
Graphs naturally introduce the inductive bias of connectivity between node pairs, which sets them apart from other data structures like images or sequential data \cite{InductiveBias}.
GNN iteratively distinguishes nodes based on their attributes, neighbors and connections. 
By leveraging the inherent graph structure of data and modeling interactions, GNNs can effectively capture the dependencies and relationships among distributed clients, leading to more accurate and robust models. 

To tackle the concerns raised earlier, a novel trusted architecture for FL is proposed.
The key contributions are summarized as follows:

\subsubsection{Privacy}
\textit{Introducing a pre-processing layer for secure aggregation.}
This layer employs HE to securely aggregate local model updates to prevent the exposure of clients’ private information.
By encrypting the updates before aggregation, only the aggregator results can be obtained by the server. 

\subsubsection{Security}
\textit{Leveraging GNNs to identify abnormal local models.}
Considering the graph structure between clients and nodes in the pre-processing layer, along with their complicated features, GNNs are employed for anomaly model identification.
To the best of our knowledge, this is the first attempt to apply GNN in the FL architecture optimization.

\subsubsection{Reliability}
\textit{Utilizing DLT for trustworthy data management and decentralized architecture.}
Using smart contracts for automatic access control and logging data exchange operations on the ledger, DLT facilitates transparent and traceable data management.
An aggregator is chosen from multiple candidates via smart contracts to execute the central server's tasks, promoting system decentralization and bolstering its resilience.

Moreover, we measure the performance in terms of abnormal model detection, global model accuracy and time overhead caused by HE.
The simulation results demonstrate that the proposed method outperforms the baseline by utilizing features of graph nodes and their connections, leading to better performance in dealing with anomaly local models. 
In addition, we analyze time consumption of system components to provide practical insights for scalability.

\section{RELATED WORK}
\label{sec:related}

Recent studies have proposed various techniques against privacy leakages, which exploit vulnerabilities to gain unauthorized access to sensitive information.
Apart from HE, commonly employed technologies include Differential Privacy (DP) \cite{DP} and secret sharing \cite{secretS}.
By adding deliberately designed noise, DP protects individual records while minimizing the impact on data aggregation \cite{DP}.
In \cite{Geyer2017}, a model aggregation algorithm was proposed to balance the trade-off between privacy loss and model performance.
The secret-sharing method distributes copies of secret messages among participants, enabling reconstruction with a certain number of participants \cite{secretS}.
A practical secret-sharing method proposed in \cite{Bonawitz2017} utilized an innovative double-mask structure to protect privacy.
Despite enhanced privacy assurances, these methods pose challenges in identifying anomalous models due to limited access to individual model updates \cite{Bonawitz2017}.

Aimed at addressing security threats in FL, considerable effort has been devoted to defense mechanisms. 
The vanilla model aggregation method is Federated Averaging (FedAvg), which performs a (weighted) average of model updates to generate global models \cite{McMahan2017}. 
Krum is a robust aggregation rule that replaces the average step with a reliable estimate of the mean \cite{Krum}. 
It selects the most trusted model as the aggregated result using majority-based and distance-based analysis.
Besides, ML-based approaches have also been extensively explored for anomaly detection, such as Multi-Layer Perceptron (MLP) \cite{MLP}.
However, these methods require knowledge of the specific data, where encryption is difficult to apply as it alters the spatial location and data distance.
Moreover, the implementations rely on a reliable third party or a centralized setting, which is not feasible in some scenarios.

Previous studies have separately explored defense mechanisms against security and privacy attacks in FL, but there is limited focus on developing combined solutions that address both threats simultaneously.
An attempt was made in \cite{Shayan2020}, but the approach to anomaly detection involves analyzing raw data, raising privacy concerns.

\section{PRELIMINARIES}
\label{sec:preliminaries}

\subsection{DLT-DDSE}
\label{subsec:DLT-DDSE}

DLT empowers traceability and immutability through decentralized peers that collectively maintain a common ledger.
Data storage in DLT is classified into on-chain and off-chain methods.
On-chain storage records model parameters directly on the ledger, while this transparency may raise privacy concerns in FL systems.
Off-chain method stores only pointers of raw data on the chain, where raw data are saved in a separate entity, such as a Distributed Data Storage Entity (DDSE) \cite{Zyskind2015}. 
DDSE can be implemented through various technologies including Distributed Hash Table (DHT), which employs hashes as keys to provide a scalable and efficient lookup mechanism.
Please refer to our previous work \cite{Ye2022} for further details on the application of DLT in FL.

In this work, we utilize DLT-DDSE to enable data exchange between clients and the central server (subsequently implemented as aggregator candidates), while ensuring transparent and immutable records.
Smart contracts are employed for automated operations, including access control and aggregator selection, thereby improving the system's trustworthiness.

\subsection{Homomorphic Encryption} 
\label{subsec:HE}

HE allows mathematical or logical operations to be performed directly on encrypted data \cite{Gilad2016}. 
$\operatorname{Enc}(m)$ denotes the encryption form of plaintext $m$.
Given encrypted values $\operatorname{Enc}(m_{1})$ and $\operatorname{Enc}(m_{2})$, homomorphic addition and multiplication operations allow for the computation of $\operatorname{Enc}(m_{1} + m_{2})$ and $\operatorname{Enc}(m_{1} * m_{2})$, respectively, without the need for decryption. 
These operations can be denoted as $\operatorname{Enc}(m_{1}) \oplus \operatorname{Enc}(m_{2}) = \operatorname{Enc}(m_{1}+m_{2})$ for homomorphic addition, and $\operatorname{Enc}(m_{1}) \otimes \operatorname{Enc}(m_{2}) = \operatorname{Enc}(m_{1} * m_{2})$ for homomorphic multiplication, where $\oplus$ and $\otimes$ represent the corresponding homomorphic operations.
%

Paillier algorithm is a public-key cryptographic algorithm based on composite residuosity classes, where encryption and decryption use modular exponentiation \cite{Paillier1999}.
It permits homomorphic addition and multiplication of ciphertext by plaintext numbers.
A detailed description can be found in \cite{Paillier1999}.

In this work, HE is used for model aggregation and linear projection, and Paillier's properties meet these demands.
On the other hand, HE schemes with homomorphic multiplication necessitate a more complex operation of noise removal after finite multiplication, leading to a substantial communication overhead \cite{Gilad2016}.
Given these factors, Paillier is selected for the encryption scheme in the proposed architecture.

\subsection{Graph Neural Network}
GNNs have gained considerable interest for their proficiency in processing and analyzing graph-structured data \cite{Scarselli2008}.
Normally, a graph consists of nodes and edges that indicate the connectivity of the node pairs. 
A bipartite graph is a specific model in graph theory, in which nodes are divided into two disjoint sets with edges connecting only nodes in different sets. 

Message passing between nodes through edges is the core mechanism of GNNs, capturing node relations and interactions. 
A GNN comprises multiple layers, with each layer creating and aggregating the message from neighboring nodes to update the representation for each node. 
This process is repeated across all layers, ultimately generating a rich, high-level representation of the nodes or the whole graph. 
GNNs have been widely applied in diverse domains, including node classification, where node embeddings are utilized and a classification head (typically an MLP) is employed.


In the proposed architecture, model features, pre-processing layer properties and their connections are essential to assess model quality.
These elements structurally make up a bipartite graph, which lacks a fixed form and is not amenable to distance measurements.
Therefore, GNN serves as an effective tool to address this challenge.

\section{ARCHITECTURE DESIGN}
\label{sec:architecture}

This section outlines the proposed trusted architecture for supporting FL, which aims to enhance both privacy and security.
It begins with an introduction of system components and threat assumption, then proceeds to detail the proposed approach, including the algorithm presentation.

\begin{figure}[t!]
    \begin{center}
    \includegraphics[width=\columnwidth]{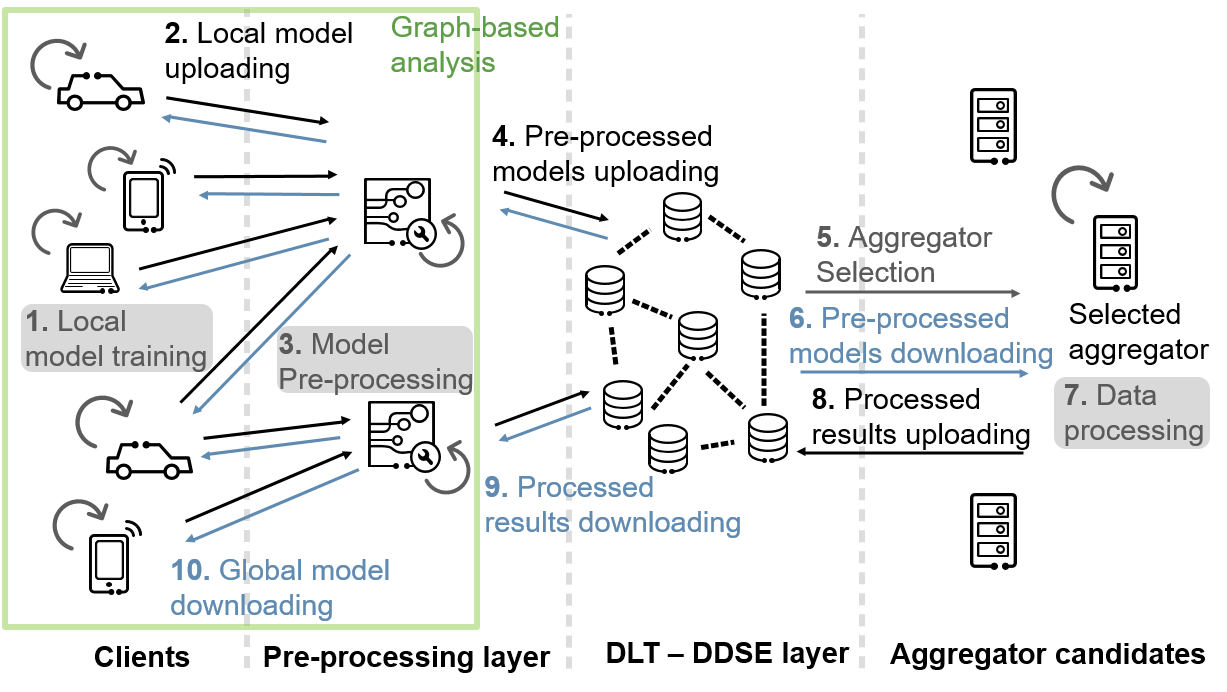}
    \caption{A trusted architecture for FL support with graph-based analysis (shown in the green box and implemented in Step 7)}
    \label{fig:architecture}
    \end{center}
\end{figure}

\subsection{System Components and Threat Assumption}
The proposed architecture consists of four components and the corresponding interactions, as illustrated in Fig.~\ref{fig:architecture}.

Clients in the proposed system download the global model, train it based on their own datasets and upload the obtained models.
Clients could be, but are not limited to, user equipment or IoT devices capable of data collection and model training, such as smartphones and cars.

The pre-processing layer performs model pre-processing before storage on the DLT-DDSE platform.
By leveraging Paillier's homomorphism, this layer implements secure aggregation and model projection with dimensionality reduction, maintaining data similarity while preventing the recovery of original data. 
Thus, subsequent detection of anomaly models is facilitated in a secure and privacy-preserving manner.

The DLT-DDSE layer is utilized for data management and aggregator selection.
All the data access records are stored as transactions in an immutable ledger.
To upload or download data, entities require authorization from DLT and then proceed accordingly to DDSE where the raw data is stored, with DDSE confirming consistency with DLT.
In addition to data management, DLT employs smart contracts for selecting aggregators, such as by random or round robin, or based on the hash of previous blocks.
The selected aggregator assumes the central server's responsibility, thereby decentralizing the entire system and enhancing its resilience.
The specific transaction types and data interaction procedures have been clearly defined and described in our previous work \cite{Ye2022}.

The selected aggregator undertakes the task of processing uploaded models from the pre-processing layer.
To identify abnormal models, it conducts graph-based analysis based on the features of clients and nodes in the pre-processing layers and the underlying graph structure. 
Then it generates the global model for the new global epoch.

Nodes in the pre-processing layer, peers in the DLT-DDSE layer and aggregator candidates could be within mobile communication systems, such as network entities with computational capabilities. 
Alternatively, they may belong to third-party organizations, where proper registration and authentication procedures are required.

This work categorizes clients as benign or malicious, with a focus on malicious clients that launch untargeted poisoning attacks, which do not involve targeted model updated attacks or evasion attacks during inference. 
Our defense mechanism operates by filtering out malicious updates that deviate sufficiently from those of benign clients.
Other components are assumed honest but curious, following system rules but attempting to deduce sensitive information from received data.

\subsection{Proposed Approach}
\label{sec:proposed}
The architecture design, illustrated in Fig.~\ref{fig:architecture}, partitions one global epoch into ten steps.
%
For global epoch $t$, $\mathbf{C}_{t}$ denotes the set of selected clients, $\mathbf{N}$ represents the set of nodes in the pre-processing layer, and $\mathbf{C}_{n,t}$ denotes the set of clients connected to node $n$.

HE employs encryption function $\mathcal{E}_{pk}$ and decryption function $\mathcal{D}_{sk}$, with public key $pk$ and private key $sk$.
$\operatorname{Enc}(x)$ denotes the encrypted form of $x$.
Clients and aggregator candidates hold the private key for decryption and the public key for encryption, while nodes in the pre-processing layer only hold the public key.
Projection function $\mathcal{P}_{pk}$ and aggregation function $\mathcal{A}_{pk}$ are implemented through homomorphic properties without the private key, where $\mathcal{P}_{pk}$ involves flattening the model into a vector and multiplying it by a matrix, and $\mathcal{A}_{pk}$ could be achieved through FedAvg.

Test function $\mathcal{T}$ evaluates model performance on unseen data using the test dataset, yielding results based on various metrics such as accuracy, error rate and F1 score.

GNN function $\mathcal{G}$ is applied to detect anomalous models using a bipartite graph comprised of clients and nodes in the pre-processing layer, as depicted in the green box in Fig.~\ref{fig:architecture}.
The local model projection and history client selection results are taken as the client features, and semi-aggregated model performance works as the features for the pre-processing layer.

\advance\topmargin by 0.05in

\begin{algorithm}[t]
\caption{Model pre-processing for each node $n \in \mathbf{N}$}
\label{alg:preprocess}
\begin{algorithmic}[1] 
\State \textbf{Input:} Encrypted local model $\operatorname{Enc}(w_{t, c_{n}}), \forall c_{n} \in \mathbf{C}_{n,t}$
\State \textbf{Require:} Projection function $\mathcal{P}_{pk}$, Aggregation function $\mathcal{A}_{pk}$
\State \textbf{Output:} Encrypted projected vector $\operatorname{Enc}(v_{t, c_{n}})$, $\forall c_{n} \in \mathbf{C}_{n,t}$, Encrypted model semi-aggregation $\operatorname{Enc}(w_{t, n})$
\State $\operatorname{Enc}(v_{t, c_{n}}) \gets \mathcal{P}_{pk}(\operatorname{Enc}(w_{t, c_{n}}))$
\State $\operatorname{Enc}(w_{t, n}) \gets \mathcal{A}_{pk}(\operatorname{Enc}(w_{t, c_{n}}), \forall c_{n} \in \mathbf{C}_{n,t})$
\end{algorithmic}
\end{algorithm}

\begin{algorithm}[t]
\caption{Data processing}
\label{alg:process}
\begin{algorithmic}[1] 
\State \textbf{Input:} Encrypted projected vector $\operatorname{Enc}(v_{t, c_{n}})$, Encrypted model semi-aggregation $\operatorname{Enc}(w_{t, n})$, Graph connection $(n, c_{n}$), $\forall c_{n} \in \mathbf{C}_{n,t}$, $\forall n \in \mathbf{N}$, History client selection records $H(\mathbf{C})$
\State \textbf{Require:} Decryption $\mathcal{D}_{sk}$, GNN function $\mathcal{G}$, Test function $\mathcal{T}$, Aggregation function $\mathcal{A}$
\State \textbf{Output:} Anomalous model detection result $\mathbf{A}_{t}$, Client selection $\mathbf{C}_{t+1}$, Semi-aggregated model performance $\{per_{t, n}, \forall n \in \mathbf{N}\}$, Global model $w_{t+1}$
\For{$c_{n} \in \mathbf{C}_{n,t}, n \in \mathbf{N}$} 
  \State $w_{t, n} \gets \mathcal{D}_{sk}(\operatorname{Enc}(w_{t, n}))$
  \State $per_{t, n} \gets \mathcal{T}(w_{t, n})$
  \State $v_{t, c_{n}} \gets \mathcal{D}_{sk}(\operatorname{Enc}(v_{t, c_{n}}))$
\EndFor
\State $\mathbf{A}_{t}, \mathbf{C}_{t+1} \gets \mathcal{G}(v_{t, c_{n}}, H(\mathbf{C}), per_{t, n}, (n, c_{n}), \forall c_{n} \in \mathbf{C}_{n,t}$, $\forall n \in \mathbf{N})$
\State $w_{t+1} \gets \mathcal{A}(w_{t, n}, \mathbf{A}_{t}, per_{t, n}, \forall n \in \mathbf{N})$
\end{algorithmic}
\end{algorithm}

Next, we provide a detailed description of steps in Fig.~\ref{fig:architecture}.

\begin{figure*}[t]
    \centering
    \begin{subfigure}{.255\linewidth}
        \centering
        \includegraphics[width=\linewidth]{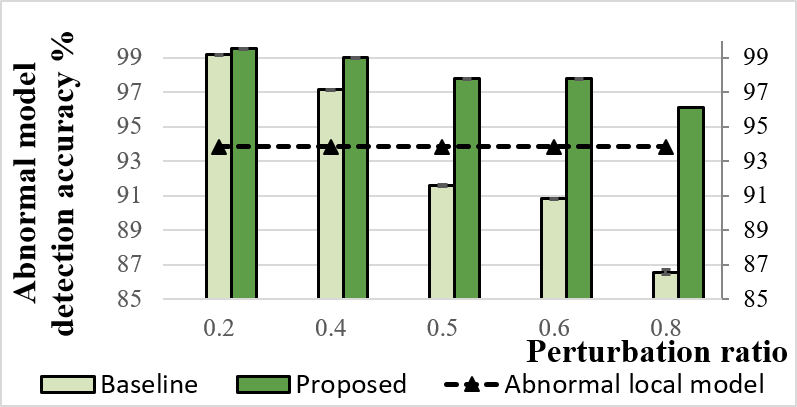}
        \caption{Perturbation ratio}
        \label{fig:subfig_pr}
    \end{subfigure}
    \begin{subfigure}{.233\linewidth}
        \centering
        \includegraphics[width=\linewidth]{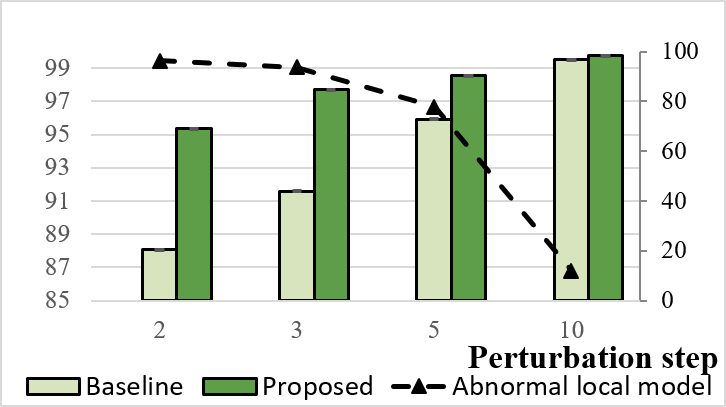}
        \caption{Perturbation step}
        \label{fig:subfig_ps}
    \end{subfigure}
    \begin{subfigure}{.233\linewidth}
        \centering
        \includegraphics[width=\linewidth]{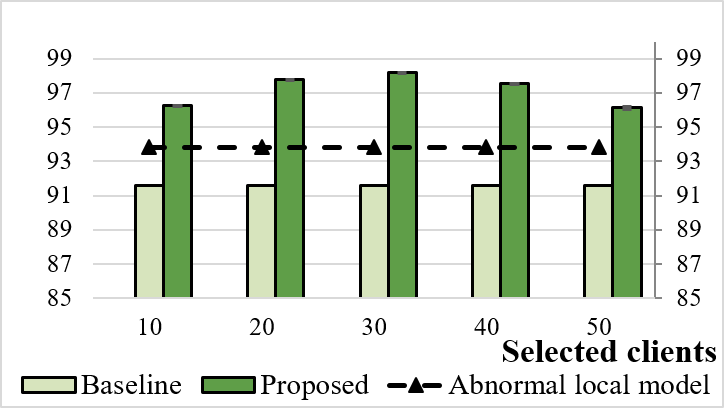}
        \caption{Selected clients}
        \label{fig:subfig_cn}
    \end{subfigure}
    \begin{subfigure}{.255\linewidth}
        \centering
        \includegraphics[width=\linewidth]{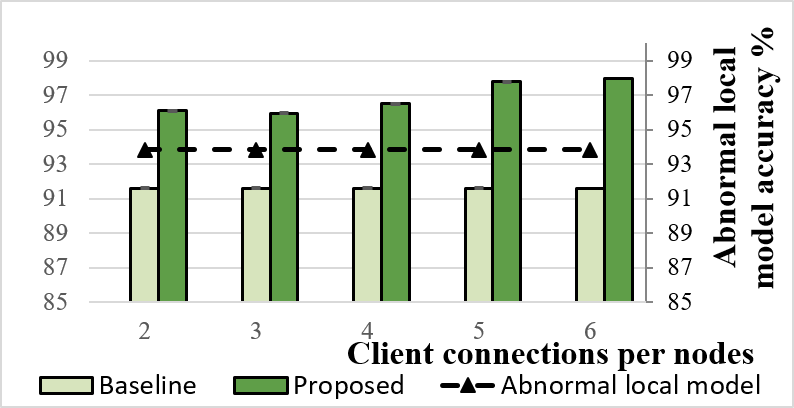}
        \caption{Client connections per node}
        \label{fig:subfig_cc}
    \end{subfigure}
    \caption{Abnormal model detection: The proposed method performs better accuracy than baseline in making use of node features and connections.}
    \label{fig:crp}
\end{figure*}

\textit{Step 1:}
Each selected client $c \in \mathbf{C}_{t}$ trains the global model $w_{t}$ on its own dataset to obtain the local model $w_{t,c}$, and encrypts it through $\mathcal{E}_{pk}$ to get $\operatorname{Enc}(w_{t, c})$.

\textit{Step 2:}
Then $\operatorname{Enc}(w_{t, c})$ is uploaded to the corresponding nodes in the pre-processing layer, where node n collects the local models from the associated clients $\mathbf{C}_{n,t}$.

\textit{Step 3:}
Nodes in this layer undertake model pre-processing, with the algorithm summarized in Alg.~\ref{alg:preprocess}.
Each node $n \in \mathbf{N}$ uses a projection function $\mathcal{P}_{pk}$ to map homomorphically encrypted local model $\operatorname{Enc}(w_{t, c_{n}})$ into a low-dimensional vector $\operatorname{Enc}(v_{t, c_{n}})$, e.g. of length 32.
Then it applies aggregation function $\mathcal{A}_{pk}$ to obtain a semi-aggregation $\operatorname{Enc}(w_{t, n})$. 

\textit{Step 4:}
The pre-processed models are passed to the DLT-DDSE layer, including $\operatorname{Enc}(v_{t, c_{n}})$, $\operatorname{Enc}(w_{t, n})$ and graph connections $(n, c_{n}$), $\forall c_{n} \in \mathbf{C}_{n,t},\forall n \in \mathbf{N}$.

\textit{Step 5:}
DLT uses the smart contract for aggregator selection. 

\textit{Step 6:}
The selected aggregator obtains pre-processed models and history client selection records $H(\mathbf{C}) = \{\mathbf{C}_{0}, \mathbf{C}_{1}, ..., \mathbf{C}_{t}\}$ from the DLT-DDSE layer.

\textit{Step 7:}
The selected aggregator performs data processing, including data decryption and analysis, as shown in Alg.~\ref{alg:process}.
The aggregator decrypts $\operatorname{Enc}(v_{t, c_{n}})$ and $\operatorname{Enc}(w_{t, n})$.
Then, the performance $per_{t, n}$ of the semi-aggregate model $w_{t,n}$ is inferred by a test function $\mathcal{T}$.
Subsequently, the aggregator takes ($v_{t, c_{n}}$, $H(\mathbf{C})$, $per_{t, n}$, $(n, c_{n})$, $\forall c_{n} \in \mathbf{C}_{n,t}$, $\forall n \in \mathbf{N}$) as the input of GNN function $\mathcal{G}$ to detect anomalous models, with the detection results denoted by $\mathbf{A}_{t}$, on the basis of which client selection $\mathbf{C}_{t+1}$ for the next global epoch $t+1$ is performed.
Semi-aggregated models with high accuracy are aggregated to form a global model $w_{t+1}$ by referring to $\mathbf{A}_{t}$.

\textit{Step 8:}
The obtained data, including $\{per_{t, n}, \forall n \in \mathbf{N}\}$, $\mathbf{A}_{t}$, $\mathbf{C}_{t+1}$ and $w_{t+1}$, are uploaded to the DLT-DDSE layer.

\textit{Step 9:}
The pre-processing layer downloads $w_{t+1}$ and $\mathbf{C}_{t+1}$, and notifies selected clients for the next global epoch $t+1$.

\textit{Step 10:}
At the end of a global epoch, selected clients $\mathbf{C}_{t+1}$ download the latest global model $w_{t+1}$.

The architecture design prevents any entity from acquiring complete knowledge of a single model and refrains from directly storing individual models on the ledger, safeguarding client information privacy. 
Simultaneously, system security is enhanced by the utilization of GNN for anomaly model detection, using graph structures and model characteristics.
All the data exchanges are immutably and transparently recorded on the DLT platform for further reference and audit.

\begin{figure*}[t]
    \centering
    \begin{subfigure}{.243\linewidth}
        \centering
        \includegraphics[width=\linewidth]{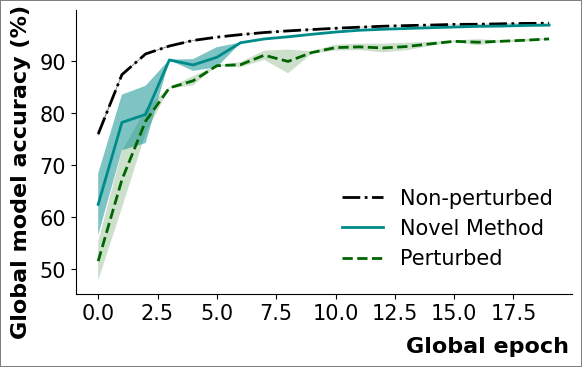}
        \caption{Basic configuration}
        \label{fig:subfig_mbs}
    \end{subfigure}
    \begin{subfigure}{.241\linewidth}
        \centering
        \includegraphics[width=\linewidth]{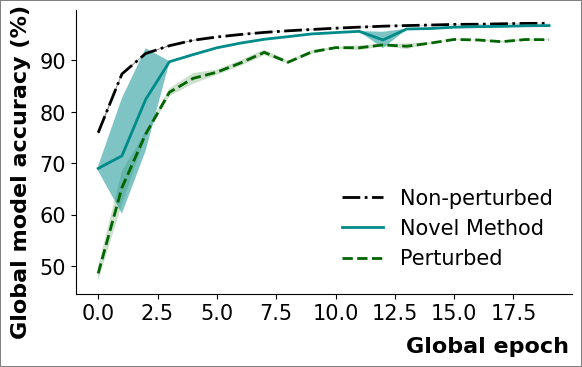}
        \caption{Perturbation ratio of 0.8}
        \label{fig:subfig_mpr}
    \end{subfigure}
    \begin{subfigure}{.241\linewidth}
        \centering
        \includegraphics[width=\linewidth]{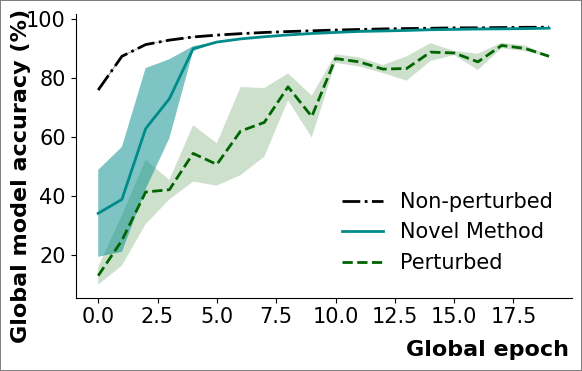}
        \caption{Perturbation step of 10}
        \label{fig:subfig_mps}
    \end{subfigure}
    \begin{subfigure}{.243\linewidth}
        \centering
        \includegraphics[width=\linewidth]{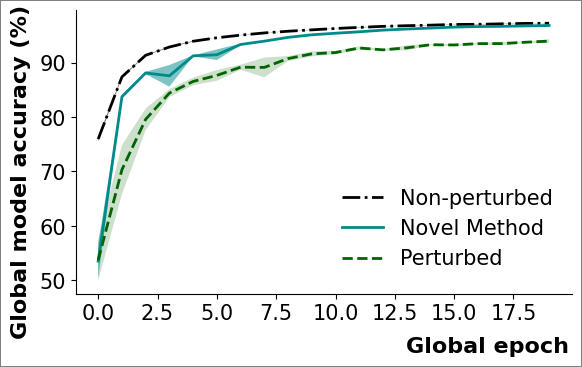}
        \caption{Client connections per node of 6}
        \label{fig:subfig_mgc}
    \end{subfigure}
    \caption{
    Global model: The novel method improves model accuracy, and the performance could be close to the non-perturbed scenario.
    }
    \label{fig:model}
\end{figure*}

\vspace{-0.3em}

\section{SIMULATION AND EVALUATION}
\label{sec:simulation}

The evaluation of the proposed trusted architecture encompasses the performance of abnormal model detection, global model accuracy and HE time overhead.
The evaluation setup is first delineated, followed by experimental results.

Regarding the performance evaluation of the DLT-DDSE, we have provided a comprehensive analysis in our prior work \cite{Ye2022}.
However, we acknowledge the significance of this component for providing trusted data management.

\subsection{Evaluation Setup}
The whole simulation is conducted on a laptop with Intel(R) Core(TM) i7-10510U CPU ({16} GB RAM), and is implemented with PyTorch 1.13.1.

\subsubsection{Dataset and model structure for FL}
MNIST dataset \cite{MNIST} is adopted for training and testing the FL models. We set {100} clients as default, each of which maintains {600} images, and trains with batch size {64}. 
For FL models, we employ a Convolutional Neural Network (CNN) model with {2} convolutional layers, both with kernel size {5}, with {32} and {64} channels respectively, followed by {2} fully connected layers of hidden size {2048}. We apply max pooling with stride {2} right after the convolutional layers.  
The employed CNN model has a total of {28,784} parameters. 

\subsubsection{Abnormal local model generation}
After the normal training process, local models are trained on a manipulated dataset with incorrect labels for several iterations, termed as \textit{perturbation steps}.

\subsubsection{Dataset and model architecture for GNN}
We use a {32}-dimensional vector as local model features, obtained through projecting the model parameters via random linear transformation matrices with a fixed seed.
For nodes in pre-processing layer, test accuracies for semi-aggregated models are taken as features.
In addition, the connection between them is taken into consideration.
For GNN, we employ a heterogeneous message passing structure consisting of {3} layers, each with {128} hidden dimensions, and dropout probability of {0.5}{}. 
Between the layers, we use ReLU activation functions for non-linearity. 
We use default Adam optimizer settings with a batch size of 64, and run each experiment for 200 epochs.

\subsubsection{Basic configuration} 
For each global epoch, the proposed system consists of 20 selected clients with 10 local epochs, 10 nodes in the pre-processing layer, each randomly connected to {5} clients.
The ratio of anomalous models, i.e., \textit{perturbation ratio}, is {0.5} with 3 perturbation steps.

\subsection{Performance of Abnormal Model Detection}
\label{subsec:clientreliability}
\subsubsection{Baseline}
The commonly used MLP model \cite{MLP} is taken as a baseline for comparison, solely leveraging projected local model vectors.
For a fair comparison, we have the same model capacity as the GNN, i.e., {3} layers with {128} hidden dimensions, accompanied by ReLU functions in-between. 

\subsubsection{Variable settings}
We examine the effect of independent variables in a controlled experimental setting.
Independent variables include \textit{perturbation ratio}, \textit{perturbation step}, number of \textit{selected client} and number of \textit{client connections per node}.
Dependent variables are MLP F1 score as the \textit{baseline} and GNN F1 score for the \textit{proposed} method, with \textit{abnormal local model accuracy} as a reference for the attack degree.

\subsubsection{Simulation result}
The comparison of our proposed method with the baseline is presented in Fig.~\ref{fig:crp}, where the primary axis displays the detection accuracy of abnormal models and the secondary axis shows the accuracy of abnormal local models.
\figurename~\ref{fig:subfig_pr} depicts the effect of increasing the ratio of malicious models, with the abnormal local model accuracy being {94}\%. 
The proposed method receives a minimal impact, with the detection accuracy remaining around 98\%.
Conversely, the baseline, which solely relies on model similarity, faces a substantial obstacle, with its accuracy dropping to 86\%.
In \figurename~\ref{fig:subfig_ps}, as perturbation steps increase, the divergence between anomalous and normal models amplifies, simplifying detection.
The novel method achieves increasing accuracy from {95}\% to {99}\%, exceeding the baseline performance from {88}\% to {99}\%. 
In \figurename~\ref{fig:subfig_cn} and \figurename~\ref{fig:subfig_cc}, changes in client numbers and connections impact the graph structure, but have no influence on the baseline with a detection accuracy of {91}\%. 
Higher connectivity enables better information aggregation, and clients that appear only once in the graph may slightly affect detection, with the lowest detection accuracy being {96}\%.
Overall, the proposed method outperforms the baseline by leveraging the features of clients and the pre-processing layer, along with their connections. 

\subsection{Performance of Global Model}
\subsubsection{Baseline}
We compare the model aggregation performance under three different scenarios: ``\textit{non-perturbed}" representing normal training without perturbation, ``\textit{perturbed}" with perturbation, and ``\textit{novel method}" with perturbation but applying the proposed method.

\subsubsection{Scenario settings}
Based on the simulation results in Section~\ref{subsec:clientreliability}, we consider application scenarios with a single change to the basic configuration, including perturbation ratio of {0.8}, perturbation step of {10}, and client connections of {6}.

\subsubsection{Simulation result}
In \figurename~\ref{fig:subfig_mbs}, the novel method achieves a quicker convergence to a higher accuracy level compared to the perturbed scenario. 
From the 10th global epoch onwards, the model accuracy obtained with the novel method is close to that in the non-perturbed condition. 
In \figurename~\ref{fig:subfig_mpr} and \figurename~\ref{fig:subfig_mps}, with an increase in perturbation ratio or step, the entire system receives more aggressive perturbations and the benefits of the novel method are amplified, with at least {8}\% higher than the perturbed condition.
The proposed architecture's performance is relatively unaffected by changes in graph structures, as shown in \figurename~\ref{fig:subfig_mgc}.
In conclusion, the novel method improves the model accuracy and stability of the FL system, and the performance could be close to the non-perturbed condition. 

\subsection{Time Consumption}


\begin{figure}[t!]
    \begin{center}
    \includegraphics[width=0.91\columnwidth]{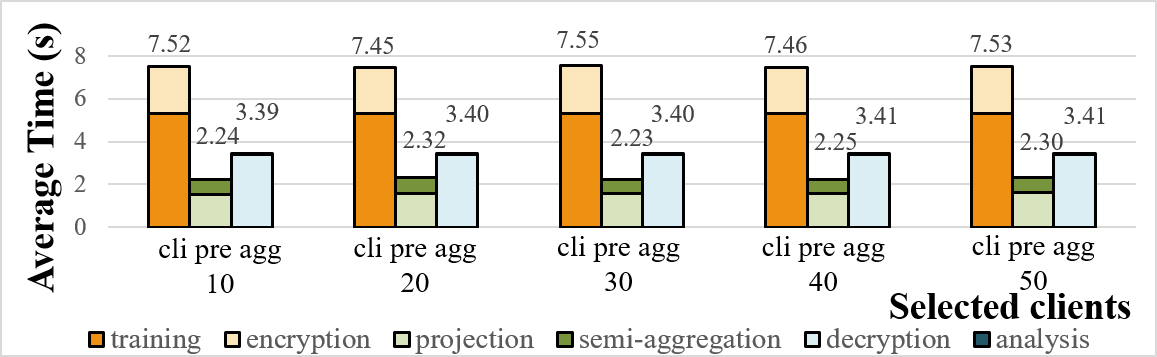}
    \caption{Average time consumption for system components per global epoch, where ``\textit{cli}" refers to clients with local model \textit{training} and \textit{encryption}, ``\textit{pre}" refers to the pre-processing layer with model \textit{projection} and \textit{semi-aggregation}, ``\textit{agg}" represents selected aggregators with data \textit{decryption} and \textit{analysis}.
    }
    \label{fig:time}
    \end{center}
\end{figure}

This subsection presents an analysis of time consumption for each system component per global epoch. 
The implemented Paillier scheme utilizes a 128-bit key length. 
We categorize the time consumption by components and investigate the impact of changes in the number of selected clients.
In Fig.~\ref{fig:time}, ``\textit{cli}" refers to clients with local model \textit{training} and local model \textit{encryption}, ``\textit{pre}" refers to the pre-processing layer with model \textit{projection} and \textit{semi-aggregation}, ``\textit{agg}" represents selected aggregators with data \textit{decryption} and data \textit{analysis}.

The majority of time is spent on training (around 40\%), encryption (around 16\%), projection (around 12\%) and decryption (around 26\%), while semi-aggregation and analysis occupy only 5\% and 0.002\% of the total time, respectively.
The increase in the client number has minimal impact on system time due to parallel training and encryption across clients, as well as parallel projection by nodes in the pre-processing layer.
Despite the additional work required for decrypting the projected vector of models with increasing clients, the impact is limited because of the vector's low dimensionality of 32.
Simulation results indicate that further processing (including pre-processing and decryption) takes longer than encryption in the client part; 
however, in practice, it does not present a performance bottleneck as it is conducted in the cloud with superior computing resources. 
Despite the overhead introduced by the Paillier scheme, the system effectively preserves privacy by safeguarding individual clients' model updates, justifying these costs. 

\section{CONCLUSION AND OUTLOOK}
\label{sec:conclusion}

This work proposes a novel trusted architecture for supporting FL.
DLT is leveraged for reliable data management and decentralized architecture.
A pre-processing layer employing Paillier's additive homomorphic encryption is incorporated to enhance client privacy protection through secure aggregation.
Considering the graph structure between clients and nodes in the pre-processing layer, GNN is leveraged to identify abnormal models, thereby improving the architecture security.
The proposed architecture enables reliable and secure collaborative machine learning in wireless communications.

This work is an exploration to apply GNN in FL architecture optimization, taking local models projection and history selection results as client features. 
In future work, we plan to incorporate geographical location and communication network characteristics, such as signal strength, into the GNN function for better prediction.

\end{document}